\documentclass[amsmath,amssymb]{revtex4}
\newcommand {\eq}{\begin{equation}}
\newcommand {\qe}{\end{equation}}

\newcommand {\bfq}{{\bf q}}

\newcommand {\bfk}{{\bf k}}
\newcommand {\bfp}{{\bf p}}
\newcommand {\h}{\frac{1}{2}}

\newcommand {\pr}{ Phys. Rev. }
\newcommand {\np}{Nucl. Phys. }

\usepackage{graphicx}% Include figure files
\usepackage{epsfig}

\begin{document}
%\baselineskip=1.2\baselineskip

\title{The pentaquark in $K^+$-d total cross section data}

\author{ W. R. Gibbs}
\affiliation{ Department of Physics, New Mexico State University \\
 Las Cruces, New Mexico 88003, USA\\}

\begin{abstract}

An analysis of $K^+$-d total cross section data is undertaken to
explore possible effects of the recently observed resonance in 
the S=+1 hadronic system with mass around 1.55 GeV. It is found that 
a structure corresponding to the resonance is visible in the data.
The width consistent with the observed deviation from background
is found to be $0.9\pm 0.3$ MeV and the mass is $1.559\pm 0.003$\ 
GeV/c$^2$ for spin-parity $\h^+$ and $1.547\pm0.002$\ GeV/c$^2$ for 
$\h^-$. The errors are one standard deviation and statistical only. 

\end{abstract}

\maketitle

\section{Introduction}

The theoretical study of the structure of a resonant state of 5 quarks
largely began with a paper by Strottman\cite{dan}.  He studied systems
in which all quarks are in an s-state and found the lowest lying state
with strangeness +1 to have spin parity $\h^-$ and a mass of around 1.7
GeV with an estimated error on the mass of 50 MeV.  If more modern
values of parameters (in particular the strange quark mass around 150
MeV instead of 279 MeV) were used, the estimate of the pentaquark mass
could be smaller.

Recently, D. Diakonov, V. Petrov, M.  Polyakov\cite{dpp} were able
to use the presumed identification of a known nucleon excited state
in the anti-decuplet to predict the mass of the isosinglet member
with strangeness +1 to be about 1.530 GeV.  They suggested that
this resonance would have a small width ($\le 15$\ MeV).  This
prediction led to a number of experimental studies which, in turn,
led to the apparent discovery
\cite{nakano,barmin,asratyan,barth,jlab1,jlab2} of a particle with
about the right mass, strangeness +1 and very probably isoscalar.
It remains to identify the spin and parity of the observed
particle, expected to be $\h^+$ from this prediction. The validity
of the soliton model used in this prediction has been
questioned\cite{cohen}.

The question is naturally raised as to why this particle was missed
in the searches that were done decades ago in the direct scattering
of positive kaons from hadronic systems. Since one must use an
incident beam of $K^+$ mesons and a neutron target (to have strangeness
+1 and isospin zero) $K^+d$ scattering is studied.

The answer to the question probably lies, at least partly, in the
very small width of the particle which appears to be emerging from
studies. The discovery experiments mentioned above are limited by
their experimental resolution so that the best they can say is that
the width is less than 9 MeV\cite{barmin}.  
Nussinov\cite{nussinov} estimated from the Fermi momentum of the
deuteron that the resonance must have a width of less than 6 MeV in
order not have been observed. Arndt et al.\cite{arndt} searched the
data base and concluded that the resonance must have a width of the
order of 1 MeV or less to have escaped notice.  

Cahn and Trilling \cite{cahn} calculated the width (from the discovery
experiment with a $K^+$ beam which observed the resonance in the charge
exchange channel\cite{barmin}) to be $0.9\pm 0.3$\ MeV. They also compared
a linear background with the same total cross section to be used later in
this work and observed a two standard deviation excess, but simply
interpreted this as an upper limit on the width of 0.8 MeV. They used a
Hulth\'en form of the deuteron wave function to calculate the Fermi
momentum correction and did not consider the effect of double scattering
or interference with the background phase shift.

The present work seeks to investigate carefully the signal to be
expected in the $K^+$d total cross section data, given the above
information.  It is found that, once the proper corrections are
taken into account to give the expected background, the signal
is indeed observed and independent determinations of the width and
mass can be obtained.

The two principal corrections necessary are the inclusion of K$^+$
double scattering and the neutron Fermi momentum in the deuteron.  
Until now, double scattering corrections for the extraction of K$^+$
amplitudes from the deuteron have been used only at higher
energies\cite{hashimoto}, even though its importance at low energies has
been known for some time\cite{humberto}.  Section \ref{double} treats
this subject.  The Fermi momentum of the nucleon in the deuteron has
been measured\cite{bern} and hence can be dealt with rather accurately.
Section \ref{fermi} treats the averaging of the amplitude over this
momentum spread. Section \ref{bps} deals with the extraction of the
background phase shifts from the proton and deuteron target data.

The studies presented here will be treated in the usual isospin
formalism and the reader is reminded of the relation between the
charge and isospin amplitudes.

\eq A_{K^+p}=T_1;\ \ A_{K^+n\rightarrow K^+n}=\h(T_1+T_0);\ \ 
A_{K^+n\rightarrow K^0p}=\h(T_1-T_0)
\qe

\section{Double Scattering\label{double}}

Double scattering has a special role in its contribution to the
total cross section for scattering from a multi-nucleon system at
low energies because of unitarity constraints in the zero energy
limit. To illustrate this point, we first look at scattering from a
simple two-body system, not very different from the deuteron.

\subsection{Low energy-weak scattering limit}

Consider double scattering from a lightly bound two body system, 
ignoring the possibilities of  spin and charge exchange. Take the 
the phase shifts to be represented by their low-energy limiting 
form 
\eq
\delta_a(k)=a k;\ \ \delta_b(k)=b k \label{phasedef}.
\qe
Since we are also considering weak scattering, $a$ and $b$ are also 
considered to be small.

In the single scattering approximation (which one might think is 
appropriate for small $a$ and $b$) and setting the bound-state form
factor to unity since we are considering the low energy limit, the 
amplitude for scattering is
\eq
f=\frac{1}{2ik}\left(e^{2iak}-1+e^{2ibk}-1\right)
\rightarrow a+b+ik(a^2+b^2)+\dots
\qe
so the elastic amplitude and cross section are, in the threshold 
limit
\eq
f_e=a+b;\ \ \sigma_e=(a+b)^2.
\qe
The integral of the elastic cross section gives the total cross 
section
\eq
\sigma_T=4\pi(a+b)^2=4\pi(a^2+b^2)+8\pi a b, \label{elastic}
\qe
since only elastic scattering is possible below the breakup 
threshold.

From the optical theorem, the total cross section is
\eq
\sigma_T=\frac{4\pi}{k}Im f(0)=4\pi (a^2+b^2).
\qe
We see that the bilinear term in $a$ and $b$ in Eq. \ref{elastic} is 
missing and the optical theorem might appear to break down. It is, 
however, the single scattering assumption which is at fault and the 
resolution of this seeming discrepancy is through double scattering.

The double scattering amplitude is given by\cite{book}
\eq
f_D(\bfk,\bfk')=\frac{1}{2\pi^2}\int \frac{d\bfq
f_b(\bfq,\bfk')f_a(\bfk,\bfq)}{q^2-k^2-i\epsilon}z\left[\h(\bfk+\bfk')
-\bfq\right]
\qe
where $\bfk$ and $\bfk'$ are the initial and final (on-shell)
momenta of the scattering meson, $z(\bfp)$ is the two-body 
form factor, and $f(\bfk,\bfq)$ and $f(\bfq,\bfk')$ are
half-off-shell basic scattering amplitudes.

\begin{center}
\begin{figure}[!tbp]
\epsfig{file=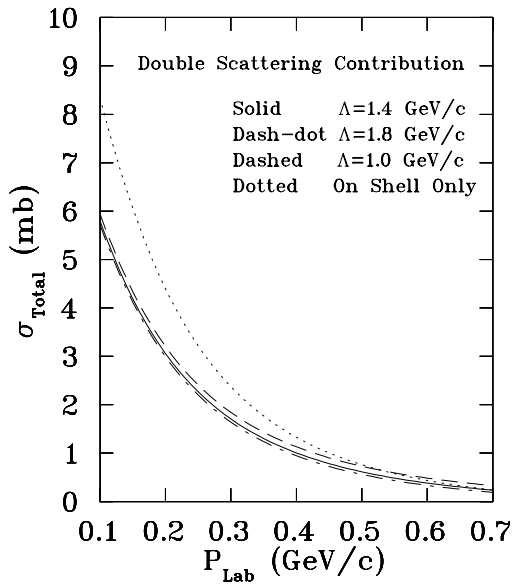,height=3in}
\caption{Double scattering contribution to $K^+$d scattering.
The common value of $\Lambda$ used for the dipole form factor 
employed in this work is around 1.4 GeV/c.}
\label{douboff}
\end{figure}
\end{center}

For s-wave scattering we write the off-shell dependence of the 
amplitude as
\eq
f(\bfq,\bfq')=f_0v(q)v(q');\ \ v(k)=1
\qe
where the form
\eq
v(q)=\left(\frac{k^2+\Lambda^2}{q^2+\Lambda^2}\right)^2
\qe
is assumed. For the limit we are considering in this section the
form is irrelevant but it is needed in the following section.

In the lowest order in $a$ and $b$, contribution of double scattering
to the forward amplitude becomes
\eq
f_D(\bfk,\bfk)\rightarrow \frac{ab}{2\pi^2}
\int \frac{d\bfq v^2(q)}{q^2-k^2-i\epsilon}z(\bfk-\bfq)
\qe
so its contribution to the total cross section is
\eq
\sigma_T^D=\frac{2ab}{k\pi}Im \int \frac{d\bfq v^2(q)}
{q^2-k^2-i\epsilon} z(\bfk-\bfq)
=\frac{2ab}{k\pi}\pi \frac{k^2}{2k}\int d\Omega \rightarrow 4\pi a b
\qe
since $z(0)=1$. With the factor of 2 which comes from the two orders
of scattering, the missing bilinear term is found. The result is 
independent of the off-shell form factor, only the on-shell 
scattering is needed.

We see that for $a$ and $b$ equal, the double scattering contributes
half of the total cross section at threshold. Even when they are
not exactly equal the contribution remains a significant fraction
of the cross section.

\subsection{Realistic case}

For the application to the present case this analysis needs several
corrections. First, the scattering lengths are small, but not so
small that corrections can be neglected. Hence, the amplitude is not 
purely real which means that the principal value of the integral gives 
a contribution and the off-shell form factor plays a role.  Second, 
charge exchange ($K^+n\rightarrow K^0p$ and its inverse) is 
possible. While single charge exchange does not lead to elastic 
scattering, so does not contribute to the forward amplitude, double 
charge exchange does. The double charge exchange is not a small 
correction, the factor $2ab$ being replaced by
\eq
2f_pf_n-f_x^2
\qe
where $f_p$, $f_n$\ and $f_x$\ are the proton, neutron and charge
exchange scattering amplitudes. The factor of 2 comes from the 2
orders of scattering, the charge exchange having only one possible
order. The minus sign is due to the isospin zero nature of the
deuteron.

Included also is the sp-wave ($S_{11}\times P_{01}$) double scattering
(on-shell only) which contributes a small negative correction at the
upper end of the momentum range in question. The notation $L_{I
2J}$, where $L$ is either $S$ or $P$, $I$ is isospin and $J$ is total
angular momentum of the partial wave, is used. The charge exchange
considerations are the same as above.

Figure \ref{douboff} shows the s wave-s wave part of the 
double scattering as used in this analysis for three typical values 
of $\Lambda$ as well as the purely on-shell contribution. While the 
differences due to the off-shell form factor are visible, the result 
is not very sensitive to the value of $\Lambda$ chosen.

\begin{center}
\begin{figure}[!tbp]
\epsfig{file=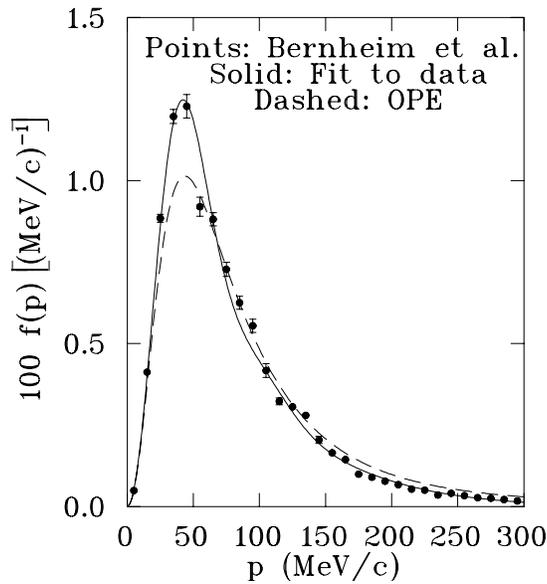,height=3in,silent=,clip=}
\caption{The measured Fermi momentum distribution compared
with the parameterization and that predicted from the 
one-pion-exchange deuteron.}
\label{bernope}
\end{figure}
\end{center}

\section{Correction for Fermi motion in the deuteron\label{fermi}}

The momentum distribution of the nucleon in the deuteron has been 
measured\cite{bern}, and these data have been parameterized\cite{dg}
as
\eq
f(p)=1.7716\times 10^{-5}p^2\left\{e^{-0.00063p^2}
+0.201e^{-0.026p}+0.0119  e^{-[(p-77.7)/38.8]^2}\right\}
\qe
with $p$ in MeV/c. Here, $f(p)$ is the probability distribution
function of the magnitude of the momentum, $p$. Notice that the $p^2$ from 
the volume element is included in $f(p)$ so that
\eq
\int_0^{\infty}dp f(p)=1.
\qe

Figure \ref{bernope} shows the data (renormalized to have integral unity)  
compared with the parameterization.  Also shown is the prediction of the
square of the momentum-space wave function of the deuteron obtained from
the solution of the Schr\"oinger equation with a one-pion-exchange
potential\cite{erc}. This deuteron wave function has been shown to
reproduce to a good approximation all of the low-energy observables
\cite{payne,ballot}.

In order to evaluate the scattering matrix in the case in which 
there is both a background and resonant phase, we take the following
form
\eq
S^b(\epsilon)=e^{2i\delta_b(\epsilon)};\ \ 
S^R(\epsilon)=\frac{\epsilon-M-i\frac{\Gamma}{2}}{\epsilon-M+
i\frac{\Gamma}{2}}=e^{2i\delta_R(\epsilon)};\ \ \delta_R(\epsilon)
=\tan^{-1}\frac{\Gamma}{2(M-\epsilon)} 
\qe
where $\epsilon=\sqrt{s}$ and $S^b(\epsilon)$ and $S^R(\epsilon)$
are the background and resonant forms of the S-matrix.

The total S-matrix including both the background and the resonance
is written as
\eq
S(\epsilon)=S^b(\epsilon)S^R(\epsilon). \label{prod}
\qe
While this the form  standardly used, a discussion of the 
representation of the S-matrix in a product form may be found in 
Ref. \cite{fluct}. 

For a kaon with momentum $\bfk$ incident on a neutron in the
deuteron with Fermi momentum $\bfp$, the square of the invariant mass
of the kaon-neutron system is given by
\eq
s=(\sqrt{\mu^2+k^2}+\sqrt{m^2+p^2})^2-(\bfk+\bfp)^2
=\mu^2+m^2+2\sqrt{\mu^2+k^2}\sqrt{m^2+p^2}-2\bfk\cdot\bfp
\qe
where $\mu$ and $m$ are the kaon and neutron masses. Due to axial 
symmetry, $s$ is independent of the azimuthal angle and 
$\bfk\cdot\bfp\equiv kpx$. 

For the case of a given isospin, $I$, and only s- and p-waves,
we may write the total cross section as
\eq
\sigma_I(\epsilon)=\frac{2\pi}{k_{cm}^2}Re 
[(1-S_{Is_{\h}})+(1-S_{Ip_{\h}})+2(1-S_{Ip_{\frac{3}{2}}})]
\equiv\frac{2\pi}{k_{cm}^2}g_I(\epsilon)
\label{sigofs}\qe

The average over the Fermi momentum distribution will give the 
observed cross section
\eq
<\sigma_I(k)>=\frac{\pi}{k_{cm}^2}\int_{-1}^1 dx \int_0^{\infty} dp 
f(p) 
g_I[\epsilon(k,x,p)]
\qe
The slowly varying factor $\frac{1}{k_{cm}^2}$ has been factored out.
\begin{figure}[b!tp]
\epsfig{file=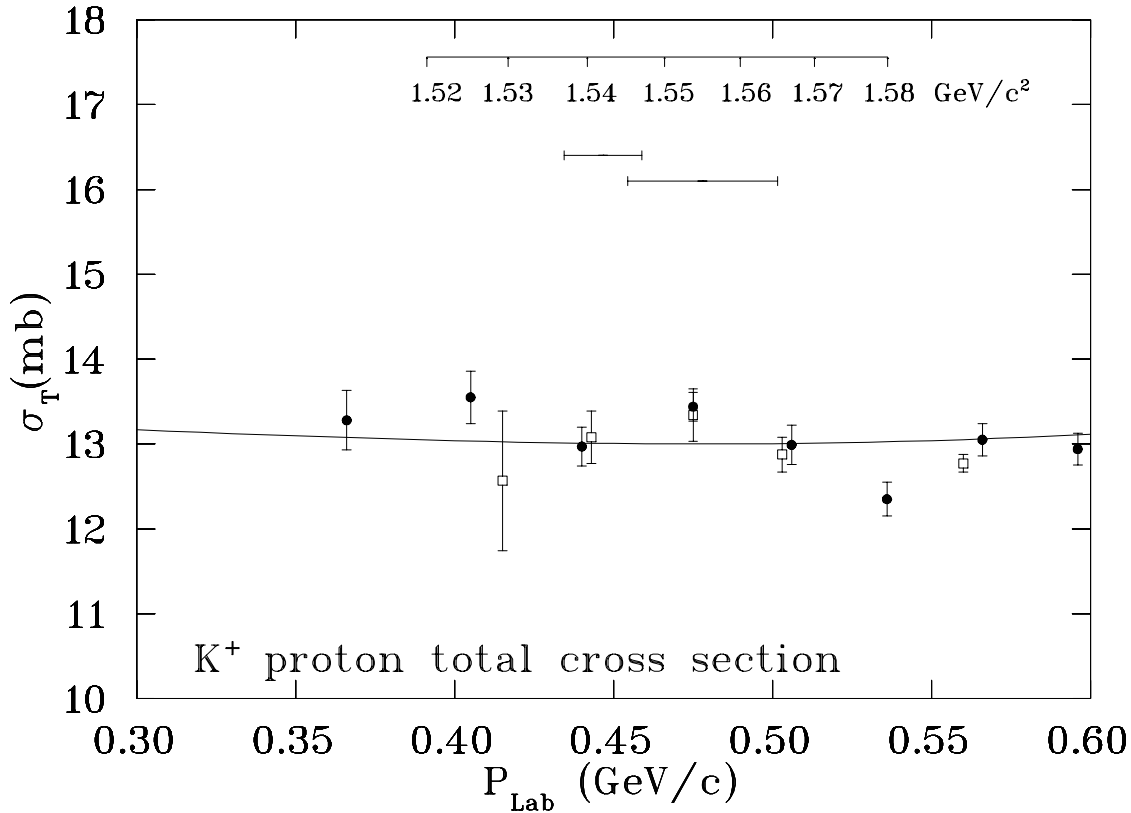,height=4in}
\caption{$K^+$ proton total cross section compared with the fit
used here. The solid points are from Bowen et 
al.\protect{\cite{bowen}} and the open points are from Carroll et 
al.\protect{\cite{carroll}}} \label{ptotalmy}
\end{figure}

It is assumed that only one term will be resonant in Eq. \ref{sigofs}. 
Its contribution to the total cross section will be
\eq
\frac{2\pi}{k_{cm}^2}\{1-\cos  
2[\delta_b(\epsilon)+\delta_R(\epsilon)]\}
\qe
which will be zero for some value of $\epsilon$ when the sum of the
phase shifts is zero or $\pi$.  Since the background phase will
normally have a magnitude smaller than $\pi$, if it is negative,
the zero will come before the true mass (when the sum is zero) and
if it is positive the zero will come after the true mass (when the
sum is $\pi$).  Thus, for a negative background phase the visible
peak will occur at a higher energy than the mass and with a
positive background phase it will occur at a lower value.  While
the Fermi averaging will smooth this behavior so that there is no
longer a zero, a shift of the peak from the true value of the mass
remains and is increased due to the positive and negative interference
of the two phase shifts above and below the resonance..

\section{Background Phase shifts\label{bps}}

The phase shifts to be used in the analysis were obtained by 
fitting data with the corrections discussed above included with an 
eye to what has been previously obtained in the literature. We will 
need I=1 and I=0 phase shifts for s- and p-waves. The phase shifts 
are expected to be very smooth (aside from the resonance, of 
course) so that the scattering length-scattering volume 
($\delta(k)=vk^3$) forms are used in all cases.

\subsection{I=1 Phase Shifts}

The I=1 phase shifts are obtained directly from K$^+$p data.  The
total cross section is shown in Fig. \ref{ptotalmy}.  It is seen
that its value is very nearly constant.  There is no indication of
a resonance, in agreement with the determination in Ref.
\cite{barth} that the observed resonance is isoscalar.  A nearly
energy dependent cross section is in reasonable agreement with the
simple representation of the phase shifts in Eq.  \ref{phasedef}.  
The cross section calculated with this form drops slightly below
the data at the highest end of the current momentum range
indicating that a small amount of p-wave contribution is needed.  
In Ref. \cite{burnstein}, experiments measuring angular
distributions in this region were reported.  They found a very
nearly isotropic angular distribution, aside from the Coulomb peak
at forward angles.  They were able to give an estimate of p-wave
strengths, although inclusion of p-waves did not decrease the
$\chi^2$ of their fit in almost all cases.

\begin{center}
\begin{figure}[!tbp]
\epsfig{file=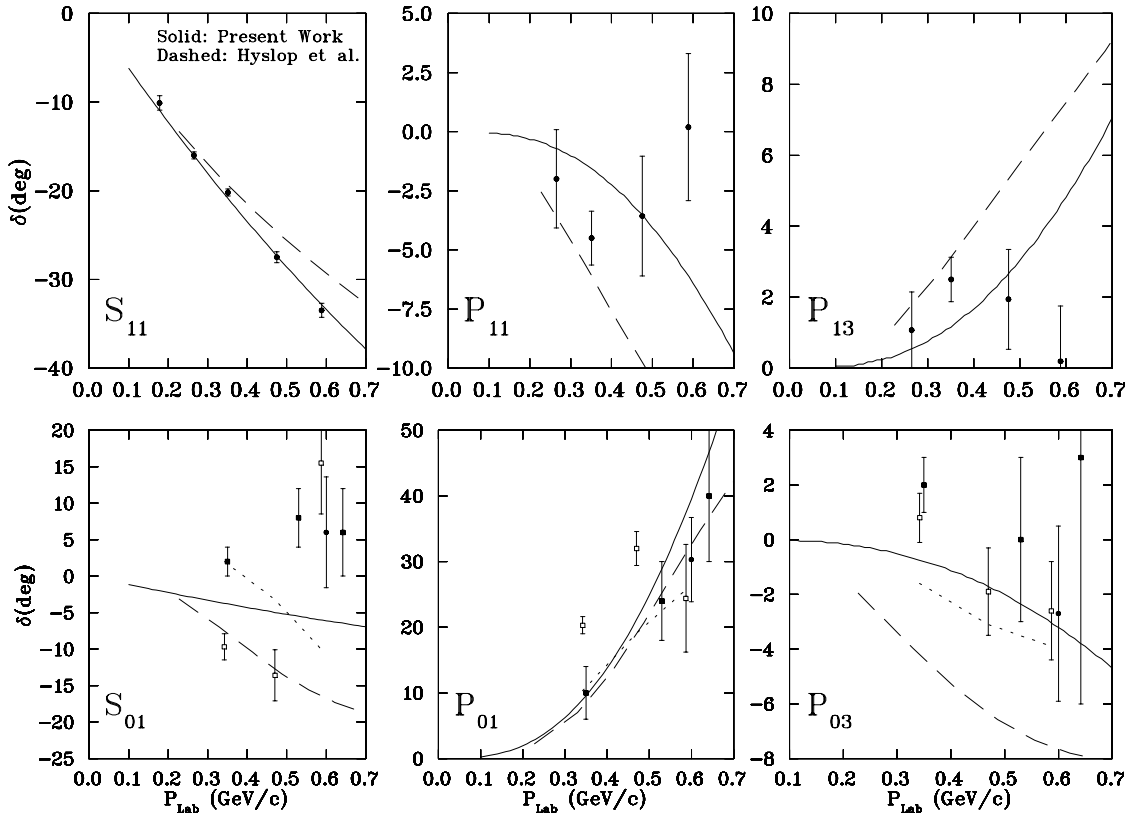,height=4in}
\caption{Phase shifts obtained in this work from the s-wave fit (solid 
curves) compared with previously obtained values. The solid points 
for I=1 are from Ref. \cite{burnstein}. For I=0, the solid circles 
are from Ref. \protect{\cite{ray}}, the open circles are from Ref. 
\cite{glasser}, the solid squares are from Ref. \cite{stenger} and the 
dotted curve is from the ``C'' fit by Ref. \protect{\cite{bgrt}}. }
\label{phases}
\end{figure}
\end{center}

The results of the present analysis are shown in the top three
panels of Fig. \ref{phases}. The solid lines give the phase shifts 
used here. The scattering lengths and volumes are given in Table I. 
The s-wave phase shifts agree very well with Ref. \cite{burnstein} 
whose  points are shown.  Also shown are the results from the 
analysis of Hyslop et al. \cite{hyslop}.

In summary, the I=1 s-wave scattering phase shifts are very well
determined in this energy range and the p-waves, although poorly
determined, are small. They have often been entirely neglected in
previous analyses.  Aside from the s-wave scattering length,
important for double scattering, the only part needed from the I=1
phases is the proton total cross section which could be taken
directly as a parameterization of the data.

\subsection{I=0 phase shifts}

The I=0 phase shifts must be inferred from analysis of scattering
from the deuteron, hence are sensitive to the corrections introduced
in Sections \ref{double} and \ref{fermi}.  The values determined
here, and a summary of previous values, are shown in the lower 
panels of Fig. \ref{phases}.

The most relevant data are the total cross sections \cite{bowen,
carroll,krauss} and the charge exchange differential cross
sections\cite{glasser}. The polarization data of Ray et
al.\cite{ray} permit the determination of the sign of the p-wave
phase shift. The data of Stenger et al.\cite{stenger} were taken
from angular distributions in a bubble chamber and are not as 
sensitive to the double scattering correction as the total cross 
section measurements.

For the total cross section data the eight points of Bowen et
al.\cite{bowen} in this momentum range are the most accurate in terms of
individual errors with a precision of 1-2\%. The Carroll et
al.\cite{carroll} data are slightly less precise.  While the Krauss et
al.\cite{krauss} data have larger error bars, they were taken with a view
to obtaining the ratio to other nuclei and hence the normalization was a
more important consideration.  Comparing the three data sets it is seen
that there is a normalization discrepancy among them. In order to bring
the normalization into agreement with the Krauss et al. data,  without 
changing the shape, the Bowen data were renormalized by 1.06. This has
almost no effect on the determination of the mass and width of the 
structure obtained later but does affect the value of the I=0 s-wave
phase shift.

\begin{center}
\begin{figure}[!tbp]
\epsfig{file=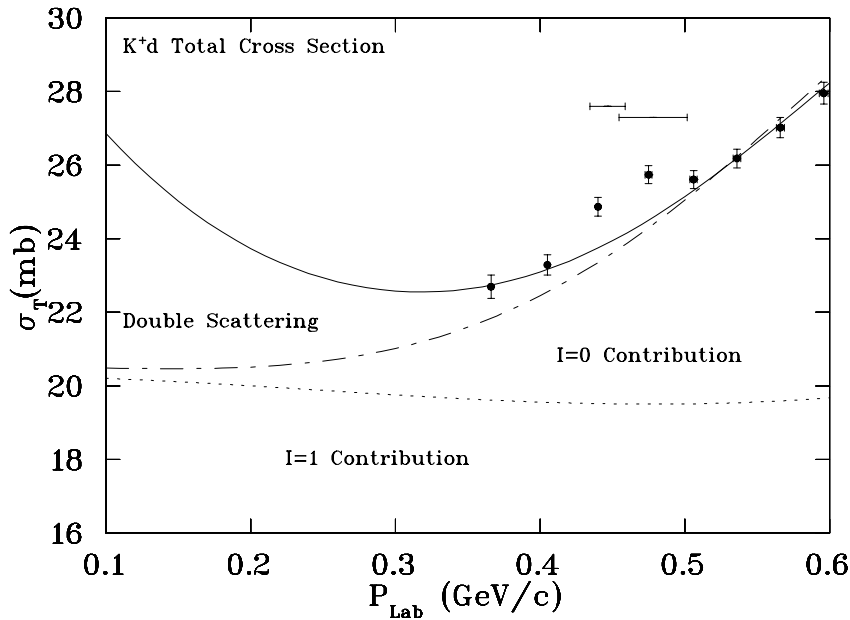,height=3in}
\caption{Background fit used in this work. The highest and lowest
momentum points were used. The solid points are from Ref.
\protect{\cite{bowen}}. The dotted curve is only the I=1 contribution,
the dash-dot curve includes that plus the I=0 contribution and the
solid curve includes the double scattering as well. The horizontal bars 
indicate the range of masses from references \cite{jlab1} and 
\cite{jlab2}. The ranges from all of the experiments can be found on figure 
7.}\label{newtots}
\end{figure}
\end{center}

The dominant I=0 phase shift is in the $P_{01}$ partial wave.
It is reasonably well determined from the total cross section data 
at the upper end of the range considered here.  

The single-energy values of the $S_{01}$ phase shift determined
previously are scattered. Note that several of them (see Fig.
\ref{phases} lower left panel) are zero or positive while others are 
significantly negative.

To determine the phase shifts to be used here, the double
scattering and Fermi corrections were applied to proposed phase
shifts derived from scattering lengths and volumes and then these
scattering lengths and volumes were adjusted to fit the data at the
high (top three points) and low (lowest two points) ends of the
data set, avoiding the intermediate region where the resonance is
expected.  The Fermi correction was applied only to the
scattering-volume form of the $P_{01}$ wave since the $P_{03}$
phase shift is very small and the $S_{01}$ wave gives an
energy-independent cross section and hence is not affected by Fermi
averaging. The result and the various contributions are shown in
Fig. \ref{newtots}.  The solid curve is very similar to that obtained
by Garcilazo from a fully relativistic Faddeev calculation\cite{humberto}.

The effect of the double scattering is to raise the cross section
at low momenta, lessening the contribution from the s-wave.  The
value of the scattering length used here is $a_0=-0.06$ fm ($0.00$
in the case of the p-wave fit, see below) so it is nearly zero, more in
agreement with the single-energy values mentioned above. 

\begin{table}[hb]
$$
\begin{array}{|c|c|c|c|}
\hline
&S_{\frac{1}{2}}\ {\rm (fm)}&P_{\frac{1}{2}}\ ({\rm 
fm}^3)&P_{\frac{3}{2}}\ ({\rm fm}^3)\\
\hline
I=1&-0.328&-0.02&0.015\\
\hline
I=0&-0.06(0.00)&0.123(0.127)&0.015\\
\hline
\end{array}
$$
\caption{Scattering lengths and volumes for the s-wave fit. The p-wave fit 
values are in parentheses. Only the $S_{01}$ and $P_{01}$ values are 
different.}
\end{table}

The $P_{01}$ partial wave obtained here agrees reasonably well with
Hyslop et al.\cite{hyslop}. The single energy values are fairly
scattered.  The $P_{03}$ partial wave is not well determined but is
very small.

\begin{center}
\begin{figure}[!tbp]
\epsfig{file=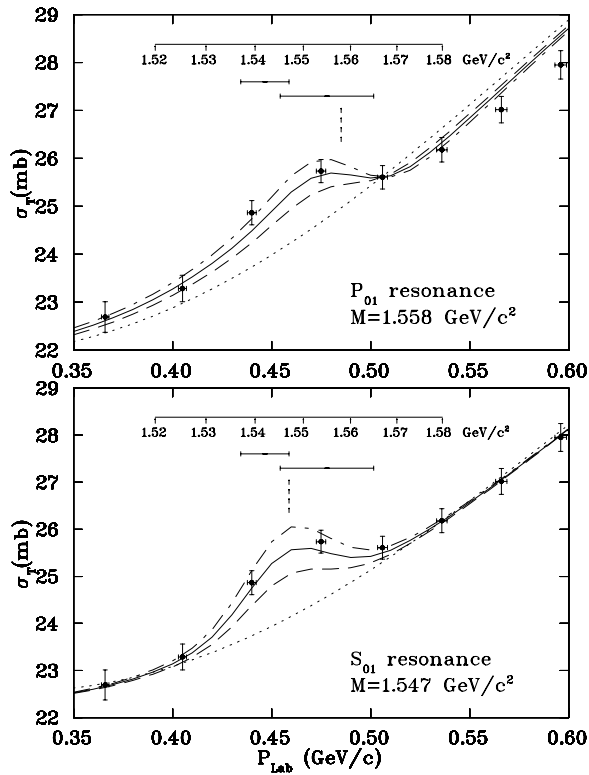,height=4in}
\caption{Comparison of Fermi motion corrected resonances with the data.
The dash-dot, solid and dashed curves correspond to widths of
1.2, 0.9 and 0.6 MeV respectively. The dotted curve is the 
background fit. The horizontal bars are the same as in 
Fig. \ref{newtots}. The vertical dashed line in each panel shows
the input value of the mass. For the s-wave resonance the 
theoretical peak occurs almost at this value while, for the p-wave 
resonance, there is a noticeable change due to the fact that the
background phase shift is considerably larger.}
\label{dtotsp}
\end{figure}
\end{center}

\begin{center}
\begin{figure}[!tbp]
\epsfig{file=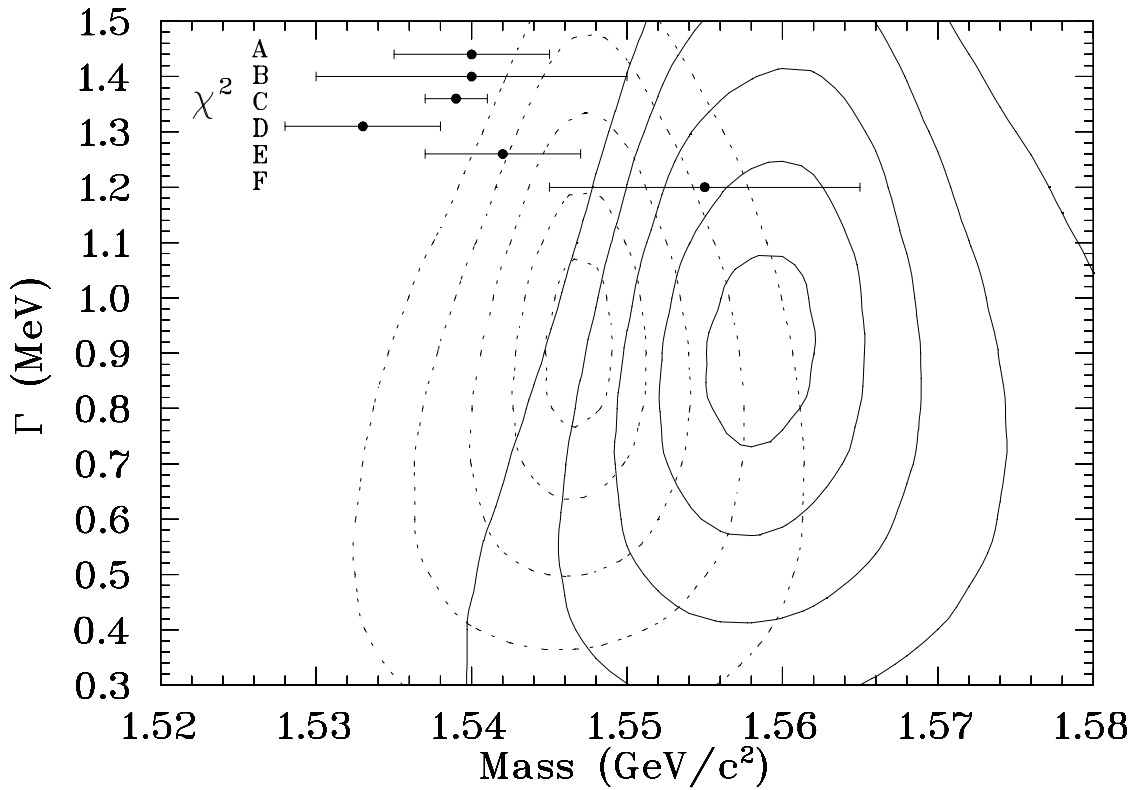,height=4in}
\caption{$\chi^2$ contour plots as a function of $\Gamma$ and
mass. The inner contour corresponds to 1 standard deviation, the
next to 2 standard deviations etc. The solid curves are for the $\h^+$ 
case (P$_{01}$ partial wave) and the dashed curves are for 
$\h^-$\ (S$_{01}$ partial wave). The points at the top correspond
to mass estimates given by the 6 discovery experiments cited in the
introduction (A-F correspond to Refs. 
\protect{\cite{nakano}-\cite{jlab2})}. The vertical placement of these 
points has no significance.}
\label{contoursp}
\end{figure}
\end{center}

\section{Results}

It can be seen that the expected cross section obtained in the
previous section (the solid curve in figure \ref{newtots}) falls
well below (5 $\sigma$) the data in the region where the resonance has
been observed in Ref. 3-8, indicating the existence of a possible
resonance effect.

The calculation of the cross section for the expected resonance is
now made as a function of a) partial wave in which is should
appear, b) width assumed and c) mass assumed.  Only partial waves
$S_{01}$ and $P_{01}$ are considered. When the $P_{01}$ partial wave was
calculated, the fit to the background had to be redone so that the high
and low momentum points were fit with the resonance since the effect of
the interference extends much further. A calculation for the
$P_{03}$ partial wave would give essentially the same result as for
the $S_{01}$ wave with a factor of 2 smaller width. If the particle
is indeed the one predicted in Ref. \cite{dpp}, it should be seen
in the $P_{01}$ partial wave although there is no a priori reason
why a particle in the $S_{01}$ state could not exist. Indeed, it
might be identified with the one predicted in Ref. \cite{dan}. 

Figure \ref{dtotsp} shows results for $\Gamma$ around the value of
0.9 MeV expected\cite{cahn}.  The mass assumed is given on the
figure and marked with the vertical dashed line.  It is seen that
the experimental deviation from the background curve is in good
agreement with the expectation for the mass chosen.  The two
assumptions for partial wave lead to equally good fits so, from
these considerations, one cannot distinguish between them with the
present data.

We see that the peak of the case for $P_{01}$ is shifted to lower
values of the mass than the input value while for the $S_{01}$
partial wave there is essentially no shift (since the phase shift is
very small).  In principle, this effect might be used as a method to
distinguish between the partial waves if an accurate determination
of the mass is made by other means. The error bands for the mass for
two of the most recent experiments (Ref. \cite{jlab1} and \cite{jlab2})
are given in the figure.

One can now determine the global best fit parameters for width and mass.
Figure \ref{contoursp} shows $\chi^2$ contour plots for the $S_{01}$
(dotted) and $P_{01}$ (solid) partial waves.  The $\chi^2$ values were
calculated by comparison of the theoretical curves (similar to those shown
in Fig. 6) with the eight points of Bowen et al.\cite{bowen}. The input
values of the mass and width were varied as shown on the axes of Fig. 7.
The inner curve in each case is one standard deviation from the minimum
(center), the next concentric curve is two standard deviations etc.  
The values and one sigma errors are read directly from the figure.
Using the full range of variation in the two fits a single value of
$\Gamma$ can be obtained as $0.9\pm 0.2$ MeV, in agreement with Cahn and
Trilling\cite{cahn}. It has been assumed that the
background fit previously is correct, i.e. it was not allow to vary.
By adjusting the background to pass through the extremes of the error 
bars, a shift of 0.2 MeV was seen. Combining the two uncertainties in
quadrature the value $\Gamma=0.9\pm 0.3$\ MeV is obtained.

The masses obtained are $1.559 \pm 0.003$ GeV/c$^2$ ($P_{01}$)and $1.547
\pm 0.002$ GeV/c$^2$ ($S_{01}$) for the two cases, where the errors are
one standard deviation only. The change in the background has negligible
effect on the masses or their errors. Aside from the statistical errors
quoted, systematic errors in the experiment (or the analysis) will
contribute as well. For example, the beam momentum was used as given.  To
move the mass from 1.547 MeV/c$^2$ (obtained for s-wave scattering) to the
nominal value of the mass obtained from the discovery experiments, 1.540
MeV/c$^2$, would require a reduction of beam momentum of 3.7\%. To move
1.559 MeV/c$^2$ (the value for the p-wave scattering) to 1.540 MeV/c$^2$
requires a reduction of 9.2\%.

Estimates for the mass from the discovery experiments cited in the
introduction are also given in Fig. \ref{contoursp}. The letters A-F
correspond to references \cite{nakano,barmin,asratyan,barth,jlab1,jlab2}
in order. We see that there are some differences in the mass
determinations.

\section{Conclusion}

It has been seen that the strangeness +1 resonance recently observed
in several experiments can also be seen in $K^+$d total cross
section measurements.  The $\chi^2$ contours given in Fig.
\ref{contoursp} indicate that the value of the background phase
influences the mass extracted to a considerable extent.  Hence, it
may be possible to infer the parity of the state from a comparison
of mass values.

The question of the parity is a very important one.  Both Karliner and
Lipkin \cite{karliner} and Jaffe and Wilczek\cite{jaffe} have proposed
models in which the small width can be explained by the partitioning of
the structure into two clusters which move relative to each other in a
p-wave, requiring an overall positive parity.  On the other hand, the
lowest lying states are most often those with all constituents in the
s-wave such as treated by Strottman\cite{dan}. Some modern calculations
also show a theoretical preference for the negative parity state from both
QCD sum rules\cite{zhu,oka} and lattice calculations\cite{csikor,sasaki}
(although Ref.  \cite{chiu} finds a positive parity). It has also been
argued that the model of Jaffe and Wilczek should have a lower lying
negative parity state\cite{zhang}.

It is tempting to say from Fig. \ref{contoursp} that the negative parity
state (S$_{01}$ wave) is closer to the centroid of the masses determined
from other measurements and hence is favored over the $\h^+$ state. While
this may be true, it would be premature to draw that conclusion.  There is
a spread among the masses of the discovery experiments and one should
expect that the errors will be reduced as further work is done.  The most
recent experiment\cite{jlab2} had a very small error on the mass from the
statistics alone, the large error bar shown being due to possible
systematic errors. If this error were reduced (without changing the
central value) then this mass would agree with the $\h^+$ case. A second
reason for caution involves the product form of the s-matrices, Eq.
\ref{prod}. While this form is commonly used, and probably incorporates
the major part of the physics correctly, the sensitivity of the shift of
the resonance peak to this assumption merits further study.  More accurate
total cross section data would be very valuable to better establish the
background and to make a more precise determination of the mass(es).

I thank William B. Kaufmann for important contributions on several
points and a very careful reading of the manuscript.  This work was
supported by the National Science Foundation under contract
PHY-0099729.

\end{document}